\documentclass[aip, preprint
 amsmath,amssymb,
 reprint,%
]{revtex4-1}

\usepackage{graphicx}
\usepackage{dcolumn}
\usepackage{bm}

\usepackage[utf8]{inputenc}
\usepackage[T1]{fontenc}
\usepackage{mathptmx}
\usepackage{etoolbox}
\usepackage{xcolor}

\makeatletter
\def\@email#1#2{%
 \endgroup
 \patchcmd{\titleblock@produce}
  {\frontmatter@RRAPformat}
  {\frontmatter@RRAPformat{\produce@RRAP{*#1\href{mailto:#2}{#2}}}\frontmatter@RRAPformat}
  {}{}
}%
\makeatother
\begin{document}

\preprint{AIP/123-QED}

\title{Rapid and efficient wavefront correction for spatially entangled photons\\using symmetrized optimization}

\author{Kiran Bajar}
    \affiliation{Department of Nuclear and Atomic Physics, Tata Institute of Fundamental Research, 400005 Mumbai, India}
\author{Ronen Shekel}
    \affiliation{Racah Institute of Physics, The Hebrew University of Jerusalem, Jerusalem 91904, Israel}
\author{Vikas S. Bhat}
\author{Rounak Chatterjee}
    \affiliation{Department of Nuclear and Atomic Physics, Tata Institute of Fundamental Research, 400005 Mumbai, India}
\author{Yaron Bromberg}
    \affiliation{Racah Institute of Physics, The Hebrew University of Jerusalem, Jerusalem 91904, Israel}
\author{Sushil Mujumdar*}
    \affiliation{Department of Nuclear and Atomic Physics, Tata Institute of Fundamental Research, 400005 Mumbai, India}
 \email{mujumdar@tifr.res.in}

\date{\today}

\begin{abstract}
Spatial entanglement is a key resource in quantum technologies, enabling applications in quantum communication, imaging, and computation. However, propagation through complex media distorts spatial correlations, posing a challenge for practical implementations. We introduce a symmetrized genetic algorithm (sGA) for adaptive wavefront correction of spatially entangled photons, leveraging the insight that only the even-parity component of wavefront distortions affects two-photon correlations. By enforcing symmetry constraints, sGA reduces the optimization parameter space by half, leading to faster convergence and improved enhancement within finite number of generations compared to standard genetic algorithms (GA). Additionally, we establish the dependence of enhancement on the signal-to-noise ratio of the feedback signal, which is controlled by detector integration time. This technique enables correction of entanglement degradation, enhancing quantum imaging, secure quantum communication, and quantum sensing in complex environments.
\end{abstract}

\maketitle

\section{\label{sec:level1}Introduction}
Spatial entanglement \cite{Schneeloch2016, Law2004} in quantum optics  is a key resource for advanced technologies such as high dimensional quantum key distribution \cite{Sun2024,Lib2024,Erhard2020} and quantum microscopy \cite{He2023,Unternahrer2018}. In quantum key distribution, the high dimensionality of spatial entanglement offers benefits such as increased communication bandwidth and a higher tolerance for quantum bit error rates \cite{Cerf2002,HighD}. In quantum microscopy, spatial entanglement enables resolution that surpasses classical diffraction limits \cite{Boto2000}. This form of entanglement is characterized by strong transverse spatial correlations and momentum anti-correlations in photon pairs, typically generated through the nonlinear process of spontaneous parametric downconversion (SPDC) \cite{Couteau2018,Karan2020}.

However, spatial correlations in entangled photon pairs are highly susceptible to degradation when the photons propagate through complex media, such as turbulent atmospheres, biological tissues, or multimode fibers \cite{Defienne2016}. This scrambling of correlations significantly impacts the practical performance and reliability of quantum technologies relying on spatial entanglement. The effects of disorder on spatially entangled photon pairs have been studied extensively \cite{2pspeckles,Peeters2010,Lib2022,Beenakker2009}. Compensating for these effects is crucial for enabling practical applications \cite{Valencia2020}. Adaptive optics is used as a tool for restoring quantum correlations, but the inherently weak quantum signal poses a significant challenge. As a remedy, researchers often employ an auxiliary classical beam, carrying an identical wavelength and spatial mode as the quantum signal, for wavefront correction \cite{Devaux2023,Courme2023}. Although effective, this method requires precise spatial alignment between the classical and quantum beams, introducing experimental complexity. Moreover, real-time correction is difficult because separating the classical beam from the quantum signal is challenging. An alternative approach leverages the pump beam for wavefront correction \cite{realtime,Shekel2021}. When a diffuser is placed in the near field of the SPDC crystal, the two-photon speckle pattern matches the speckle pattern of the pump beam, enabling real-time restoration of two-photon correlations through feedback and control of the pump beam. In contrast, when the diffuser is positioned in the far field of the SPDC crystal, it is helpful to decompose the diffuser function in odd- and even-parity components \cite{Bajar2024, Black2019}. In this case, only the even-parity component of the diffuser contributes to wavefront distortion, while the odd-parity component does not affect the two-photon correlation \cite{Bajar2024}. This insight enables more efficient wavefront correction, as adaptive optics can focus exclusively on compensating the even-parity distortions.

In this article, we demonstrate the restoration of two-photon correlations for high-dimensional spatially entangled photons without the need for an auxiliary classical beam. By leveraging the premise that only the even-parity component of the diffuser contributes to two-photon speckle in the far field, we optimize wavefront correction to target exclusively the even-parity distortions. This targeted approach reduces the wavefront correction time by up to a factor of $4$ while achieving the same level of enhancement in two-photon correlations. Furthermore, for a fixed correction time, concentrating on the symmetric (even) part of the wavefront, we demonstrate up to $38\%$ increase in the enhancement compared to correcting the entire wavefront.

\section{Theoretical background}
The two-photon wavefunction generated via collinear SPDC in terms of transverse momenta $k_1$ and $k_2$ is expressed as \cite{Schneeloch2016}:
\begin{eqnarray}
\psi(k_1,k_2) \propto e^{-(k_1+k_2)^2\sigma_+^2/{2}}e^{-(k_1-k_2)^2\sigma_-^2/{2}}|k_1\rangle |k_2\rangle
\label{eq1}
\end{eqnarray}
where, for a loosely focused beam and a thin crystal, $\sigma_+ \gg \sigma_-$.
This wavefunction indicates that the two photons are anti-correlated in transverse momentum. The Fourier transform of the wavefunction further implies that they are correlated in the transverse position.
However, introducing a lossless diffuser characterized by $A_d = e^{i\phi(k)}$, where $\phi(k)$ is a random phase function in momentum space, disrupts the spatial correlations of the photon pairs, giving rise to the phenomenon of two-photon speckle \cite{2pspeckles}.
Assuming the scale length of the disorder in momentum space is much larger than $1/\sigma_+$, the term $e^{-(k_1+k_2)^2\sigma_+^2/{2}}$ can be approximated as $\delta(k_1+k_2)$ for a loosely focused beam. Under this approximation, the two-photon speckle is described by the correlation function \cite{Bajar2024}:
\begin{eqnarray}
C(x_1,x_2)\propto \left|\int dk_1 e^{ik_1(x_1-x_2)} 
e^{i(\phi(k_1)+\phi(-k_1))}
e^{-2k_1^2\sigma_-^2}\right|^2 \nonumber\\
\label{eq2}
\end{eqnarray}
This equation reveals that only the even-parity component of the phase modulation introduced by the diffuser, $\phi(k_1)+\phi(-k_1))$,  contributes to the two-photon correlation. The odd-parity component does not influence the correlation.  

This phenomenon can be understood using Klyshko’s advanced wave picture (AWP)\cite{Zheng2024, Shekel2024}. In the AWP, one of the two detectors is conceptually replaced by a laser beam, and the nonlinear crystal is replaced by a mirror. The laser beam propagates to the mirror, reflects off it, and is subsequently detected by the second detector. In this classical analogy, the intensity measured at the detector is directly proportional to the coincidence rate observed in the two-photon experiment. For the case described above, when odd-parity phase modulation is introduced, the laser beam accumulates equal and opposite phase shifts before and after reflection from the mirror. As a result, the net phase change cancels out, meaning that the odd-parity phase modulation does not affect the detected intensity or, equivalently, the two-photon coincidence rate.

Therefore, while using adaptive optics for restoring the two-photon correlation, correcting only the even-parity contribution is sufficient. This reduces the number of independent pixels required for wavefront correction by half, thereby significantly simplifying and accelerating the optimization process.

\section{Experimental Setup}
\begin{figure}
\includegraphics[width=\linewidth]{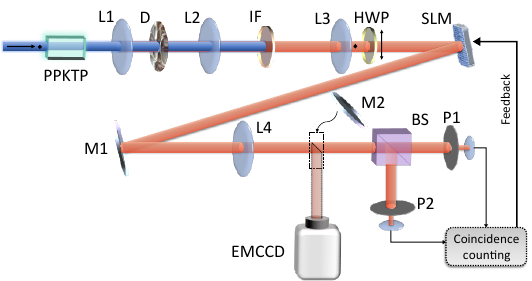}
\caption{\label{fig:1} Schematic of experimental setup. PPKTP: Periodically-poled potassium titanyl phosphate crystal, L: Lens, D: Diffuser,  IF: $810$ nm bandpass interference filter, HWP: Half Wave-plate for $810$ nm, M: Mirror, EMCCD: Electron multiplying charge-coupled device, BS: Beamsplitter, P: Pinhole. The diffuser and the SLM are positioned in the Fourier plane of the crystal plane, while the detectors (P1, P2, EMCCD) are positioned in a plane conjugate to the crystal plane.}
\end{figure}

Fig~\ref{fig:1} illustrates the schematic of the experimental setup. A Gaussian pump beam with a wavelength of $405$ nm passes through a $5$ mm long Type-0 nonlinear periodically poled potassium titanyl phosphate (PPKTP) crystal. This nonlinear crystal facilitates the generation of photon pairs through the process of SPDC, producing photons centered around $810$ nm that are spatially entangled with a Schmidt number $\approx300$.

Using lens L1, the far-field plane of the crystal's center is imaged onto a thin polymer-on-glass lossless diffuser, D, with a diffusion angle of $1^\circ$. The far-field plane is subsequently imaged onto a spatial light modulator (SLM) through a 4f optical setup comprising lenses L2 and L3. The bandpass interference filter (IF) centered at $810$ nm is placed after L2 to block the pump, allowing only near-degenerate photon pairs to pass. A half-wave plate (HWP) rotates the polarization of the entangled photons to horizontal, enabling phase modulation of the photons by the reflection-type phase-only SLM. The photon pairs are incident on the SLM at an angle of $\approx 5^\circ$. A mirror M1 further redirects the photon pairs towards the detection setup. Lens L4 performs the Fourier transform of the field at SLM, projecting the near-field image of the crystal's center onto the detector system. Two detection schemes were employed: depending on the position of mirror M2, either a pinhole pair (P1 and P2) or an EMCCD was used as the detector. The photons from the pinhole pair are directed to a Single Photon Avalanche Diode (SPAD) pair, which provides feedback to the SLM for wavefront correction, while the EMCCD captures the two-photon correlation pattern before and after correction. To block stray photons, $810$ nm bandpass filters with FWHM $10$ nm were placed in front of the pinholes and EMCCD. 

\begin{figure*}[ht!]
\includegraphics[width=\linewidth]{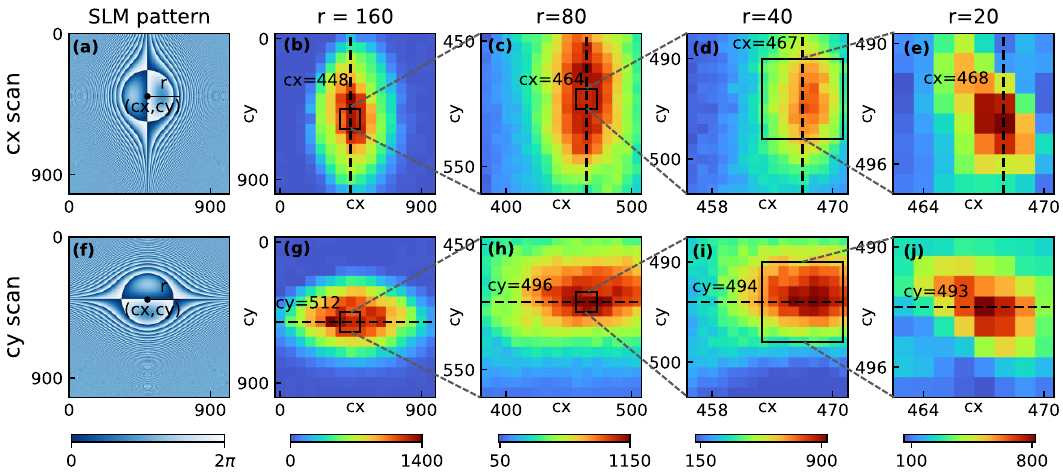}
\caption{\label{fig:2} Finding the SPDC beam center (X, Y) on the SLM. (a, e) Odd Zernike polynomials ComaX and ComaY with radius, r, and center, (cx,cy), having symmetry that preserves two-photon correlations when symmetrically placed about the beam. As a result, the correlation quality is maximized when (cx,cy) matches (X,Y). The top (bottom) panels show the correlation quality as a function of (cx,cy) for ComaX (ComaY) scans, corresponding to the determination of X (Y). (b, f) Coarse scans with r=160; regions of highest counts are zoomed in for finer scans: (c, g) r = 80; (d, h) r = 40 and (e, j) final scans with r = 20.}
\end{figure*}

A loosely focused pump beam produces photon pairs that exhibit strong transverse positional correlations in the crystal's near field \cite{Bhat2024, Bhat2025, Schneeloch2016}. However, the introduction of the diffuser, D, disrupts these correlations by deforming the two-photon wavefronts, leading to a randomized pattern in coincidence space known as two-photon speckle \cite{Beenakker2009, 2pspeckles}. To counteract this, a wavefront correction algorithm is implemented. This algorithm uses coincidence counts at spatially correlated points as feedback to the SLM, enabling the restoration of the original correlations. A beam splitter (BS) probabilistically separates the photon pairs into two paths, enabling coincidence detection. P1 and P2, are placed at correlated positions at the beam center, spatially filtering the photons. The size of the pinholes was selected by measuring the two-photon correlation width \cite{Bhat2025, Schneeloch2016}. These filtered photons are then coupled into multimode optical fibers and directed to the SPADs (not shown in Fig.~\ref{fig:1}). The photon detection events are time-tagged to compute coincidence counts, which serve as feedback for the SLM in the adaptive wavefront correction process.

 The focal lengths of the lenses used in the setup are as follows: Lens L1 comprises 3 lenses (f = $25$ mm, $75$ mm, and $25$ mm, arranged with separations equal to the sum of their respective focal lengths. The crystal's center is positioned at the front focal plane of the first lens of L1, and the diffuser, D, is placed at the back focal plane of the last lens of L1. The focal lengths of lenses L2, L3, and L4 are $50$ mm, $200$ mm, and $500$ mm, respectively.

The two-photon correlations, both before and after optimization, are measured by directing the entangled photons toward the EMCCD camera, achieved by inserting mirror M2 into the optical path. To capture the whole beam onto the EMCCD chip, the near field plane of the crystal's center is de-magnified by a factor of 3 using single-lens imaging (f = $35$ mm, not shown in the schematic). The EMCCD functions as a multi-pixel single-photon detector \cite{emccd, Hugo, MTh}. A total of $N$ frames were captured, ensuring an average photon count of  $\approx 0.1$ photons per pixel per frame to minimize photon pile-up. The electron counts recorded by the EMCCD were converted into photon numbers using a predefined thresholding technique \cite{emccd}. The captured frames were then processed through coincidence counting and background subtraction to compute the two-dimensional (2D) correlation function \cite{Bajar2024}.

For the correction of wavefront using even-parity SLM phase, it is essential to locate the center of SPDC beam, (X, Y) on the SLM. Fig.~\ref{fig:2} illustrates the process of achieving the same by implementing odd Zernike polynomials on the SLM. Specifically, ComaX and ComaY, shown in panels (a) and (e), respectively, are used. These polynomials are characterized by their radius, r, and center coordinates (cx, cy). The procedure is performed by using the experimental setup shown in Fig.~\ref{fig:1} without the diffuser, D. According to the theory, a symmetrically positioned odd Zernike polynomial in the far field of the crystal's center does not affect the two-photon correlation. This implies that when (cx, cy) coincide with (X, Y), the coincidence counts at correlated points in the conjugate plane of crystal should be maximized. We define the \textit{correlation quality} as the coincidence counts recorded by the SPADs in a given integration time and use this criterion to determine (X, Y). The top panel details the procedure for finding X, while the bottom panel focuses on Y. For each case, (cx, cy) is scanned for the respective polynomial function (ComaX for X and ComaY for Y), and the correlation quality is recorded. Panels (b) and (g) depict a coarse scan with $r = 160$ and a step size of $64$. Here, the region with the highest coincidence counts is identified and zoomed in for a finer scan, as shown in panels (c) and (h), using $r = 80$ and a step size of $8$. Further refinement is performed in the regions of maximum coincidence counts from the finer scan, as depicted in panels (d) and (i) with $r = 40$ and a step size of $1$. Each measurement in these scans had an integration time of $1$ s. The final scan, shown in panels (e) and (j), uses $r = 20$ and focuses on a cropped region to determine the beam center. For this final scan, the integration time for each (cx, cy) measurement was increased to $5$ s to improve accuracy. This shows that the center of the beam, (X, Y) is (468,493). 
This method of determining (X, Y) is free from aberrations introduced by optics after the SLM. Additionally, its sensitivity can be tuned by adjusting the value of $r$, enabling greater precision in locating the beam center, a crucial parameter for the experiment (see Appendix \ref{ap:disCenter}).

\begin{figure*}
\includegraphics[width=0.7\linewidth]{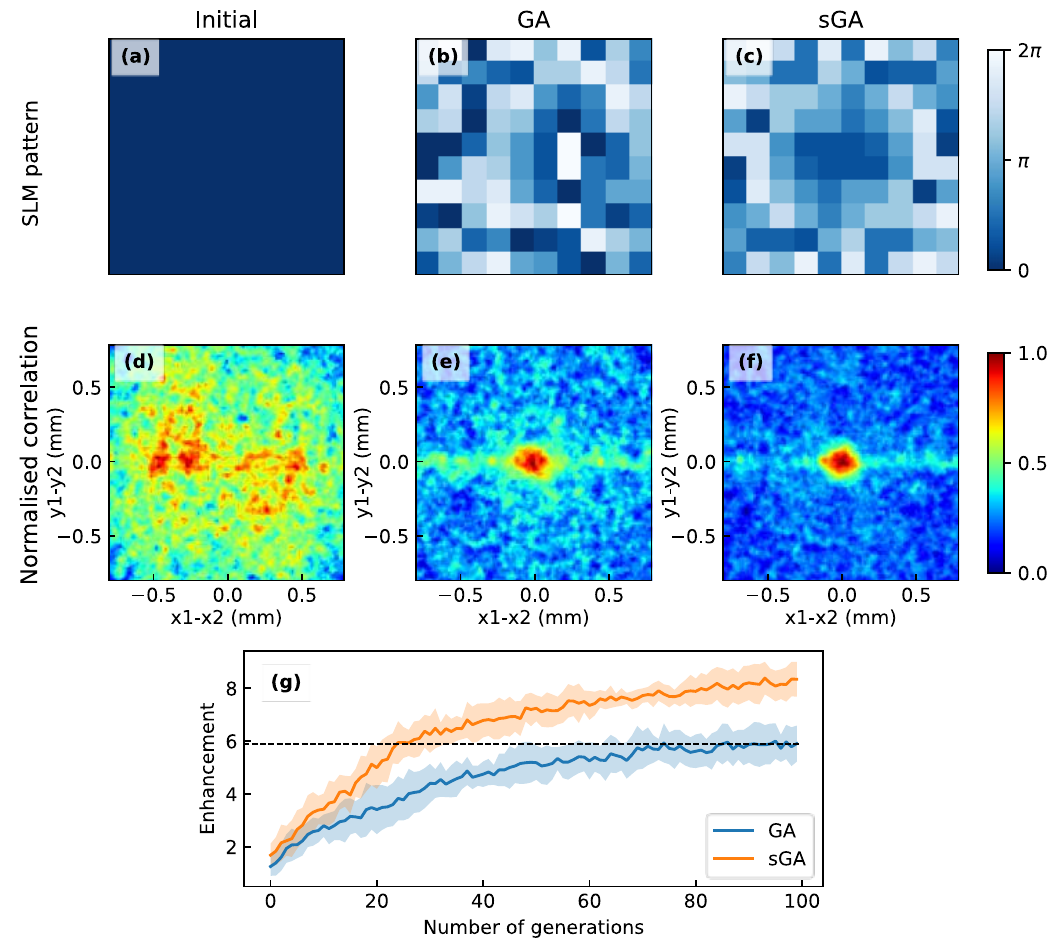}
\caption{\label{fig:3} Comparison of wavefront correction algorithms. (a–c) SLM patterns before correction, after the genetic algorithm (GA), and after sGA (even part only). (d–f) Corresponding 2D correlations recorded via EMCCD. (g) Evolution of normalized coincidence counts at correlated points over generations for GA and sGA.}
\end{figure*}

\section{Results}
The circular beam on the SLM had a diameter of approximately 80 pixels. For wavefront correction, the pixels were grouped into super-pixels, each consisting of $8\times8$ individual pixels. Within a super-pixel, the phase of all pixels is modulated simultaneously, while the phase of each super-pixel can be adjusted independently. Each super-pixel's phase was constrained to 16 equispaced discrete values ranging from 0 to $2\pi$ to reduce the search space, improving optimization efficiency. A genetic algorithm (GA) \cite{Samanta2023,Fayyaz2019,Conkey2012,Vellekoop2015} was employed for optimization. The algorithm parameters included a population size of $15$, a mutation rate of $0.1$, and the retention of three elite solutions from one generation to the next. The GA was executed for $100$ generations, with an acquisition time of $1$ s per SLM pattern for the SPADs, resulting in a total runtime of approximately $25$ minutes. 

Two optimization modes were used: GA, which corrected the entire wavefront, and sGA, which corrected only the even part of the phase front. In the sGA mode, the super-pixels of the upper half of the frame were adjusted independently, while those of the lower half were determined by enforcing inversion symmetry, i.e., $180^\circ$ rotation symmetry. 

Fig.~\ref{fig:3} presents the results of wavefront correction using the two algorithms. Panels (a-c) respectively depict the SLM patterns before correction, after correction using the GA, and after correction using the sGA. Panels (d-f) show the normalized 2D correlations corresponding to the SLM patterns in (a-c) calculated using data recoded by EMCCD. For the 2D correlation in Fig.\ref{fig:3}(d), approximately 800,000 frames were captured using the EMCCD. In contrast, for Figs.\ref{fig:3}(e) and \ref{fig:3}(f), approximately 100,000 frames were sufficient. Both algorithms successfully restored the randomized correlation, as observed in Figs.\ref{fig:3}(e) and \ref{fig:3}(f).  
We quantify the contrast of the peak using the following expression:
\begin{equation}
C = \frac{\mu(Peak)-\mu(BG)}{\sigma(BG)}
\end{equation}
where $\mu$ and $\sigma$ denote the mean and standard deviation, respectively. The $Peak$ region corresponds to the central square of $0.16~\mu$m $\times$   $0.16~\mu$m, consisting the peak and $BG$ represents the background, defined as the entire image excluding the peak region. The calculated contrast values for Fig.~\ref{fig:3}(e) and \ref{fig:3}(f) are 7.6 and 5.5, respectively. Therefore Fig.\ref{fig:3}(f) exhibits higher contrast than Fig.~\ref{fig:3}(e), indicating that sGA achieves superior performance with the same number of generations. The bright horizontal line observed in fig.~\ref{fig:3}(e),(f) is due to charge blooming in the EMCCD\cite{blooming}. For clarity this artifact has been removed in the post-analysis from fig.~\ref{fig:3}(d), where it was more prominent. Fig.~\ref{fig:3}(g) illustrates the evolution of the two algorithms by tracking enhancement across generations by using coincidence counts recorded using SPADs. We quantify the enhancement as the ratio of the highest coincidence counts in each generation to those obtained with a flat SLM at the correlated points. Since coincidence counts—and consequently, enhancement—can vary significantly across different diffuser realizations, we maintain the same diffuser realization throughout the experiment to ensure a fair comparison. Both algorithms were executed five times with different random initial configurations of the SLM while keeping the disorder realization unchanged.  The number of generations was chosen as a trade-off between enhancement and runtime. The shaded region around the solid lines represents the error bars, calculated as one standard deviation of the statistical error at each point across different runs. The comparison shows that at the end of $100$ generations, sGA outperforms GA by $38\%$. Additionally, GA reaches an enhancement of $5.9$ at the end of $100$ generations (represented by the dotted line), whereas sGA achieves the same enhancement in just $25$ generations. This demonstrates that sGA is four times faster than GA. The accelerated enhancement of sGA can be attributed to the reduced search space of the optimization algorithm. Additionally, its improved performance can be explained as follows: In GA, there is no unique solution—adding an odd-parity function to an existing solution results in another valid solution. Consequently, different parent solutions in GA may converge toward different local optima, making it difficult for the algorithm to refine the solution further. In contrast, sGA enforces a unique solution by restricting odd-parity functions, allowing the algorithm to move more efficiently toward the optimal solution.  
Importantly, the detectors need to be placed at the center of the beam to ensure a strong feedback signal. In the event of deviation of the detectors from the center, the advantages of the sGA are defeated because beam tilt, an odd function, is not allowed in sGA (see Appendix \ref{ap:ConSLM}).

Fig.~\ref{fig:4} illustrates the impact of detector integration time on wavefront correction. Fig.~\ref{fig:4}(a) presents the evolution of enhancement for three different integration times ($0.2$ s, $0.5$ s, and $1$ s) using the sGA algorithm. Each integration time was tested across five independent runs with different initial configurations of SLM for the same realization of disorder. The results indicate that enhancement improves with longer integration times. This trend can be attributed to the weak feedback signal, which is initially dominated by high noise levels due to inherent quantum fluctuations following Poisson statistics. With increased integration time, the feedback signal becomes stronger, leading to a higher signal-to-noise ratio and improved enhancement. However, this improvement comes at the cost of increased optimization time. Fig.~\ref{fig:4}(b) compares the final enhancement achieved after 100 generations for both GA and sGA. To calculate the final enhancement, coincidence counts were recorded over $50$ s for two different SLM configurations: the optimized SLM pattern after 100 generations and a flat SLM. The enhancement factor was obtained by taking the ratio of these two measurements. The error bars, represented by black lines, denote one standard deviation of the statistical error in the final enhancement. These indicate that, along with the overall increase in enhancement, the uncertainty in the final enhancement decreases with longer integration times for both algorithms.

\begin{figure}
\includegraphics[width=\linewidth]{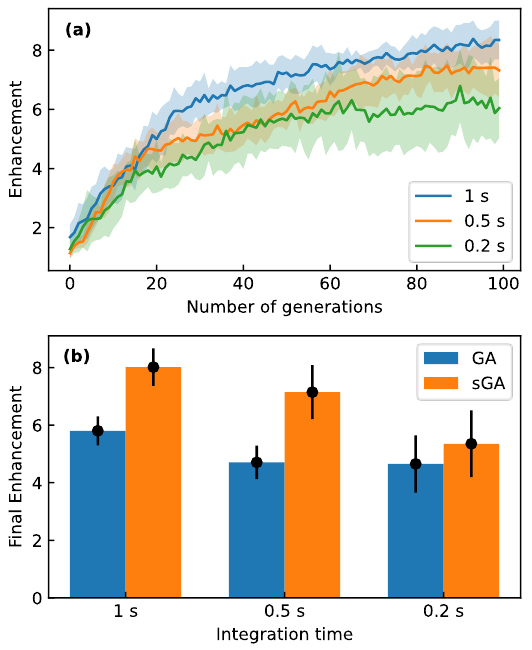}
\caption{\label{fig:4} Effect of integration time on the wavefront correction. (a) Evolution of normalized coincidence counts for sGA for three different integration times. (b) Normalized coincidence counts for GA and sGA corresponding to the 3 different integration times at the end of 100 generations.}
\end{figure}

\section{Conclusion}
We demonstrated wavefront correction for spatially entangled photons using adaptive optics without an auxiliary classical beam. By leveraging the insight that only the even-parity component of a diffuser in the far field of crystal contributes to wavefront distortion, we have presented a method to achieve faster and higher enhancement in the two-photon correlations. The crucial step of identifying the beam center falling on the SLM was implemented using an odd Zernike. Our results show that sGA outperforms the conventional GA. Further, we demonstrated that the enhancement depends on the signal-to-noise ratio of the optimization parameter, which, in turn, is influenced by the integration time of detectors. The proposed method can be combined with the strategy outlined earlier\cite{realtime}, enabling simultaneous correction of the wavefront in both the near-field and far-field planes of the SPDC crystal.  Additionally, advanced technologies such as SPAD cameras, which can collect signals from multiple correlated points, significantly improve the signal-to-noise ratio. This enhancement makes the symmetrized genetic algorithm (sGA) more effective for active wavefront correction using direct feedback from the quantum signal \cite{courme2025}, even under low-signal conditions. Together, these developments expand the applicability of our approach, making it well-suited for practical quantum optics systems, including quantum imaging and secure quantum communication.  

\begin{acknowledgments}
We sincerely thank Prof. Umakant Rapol for generously providing access to his Spatial Light Modulator (SLM) for the experiment. We thank the Department of Atomic Energy, Government of India, for funding for Project Identification No. RTI4002 under DAE OM No. 1303/1/2020/R\&D-II/DAE/5567, Ministry of Science and Technology, India. The authors declare no conflict of interest.
\end{acknowledgments}

\section*{Data Availability Statement}

The data that support the findings of this study are available from the corresponding author upon reasonable request.

\appendix

\section{Effect of displacement of optimization center (cx, cy) on SLM}\label{ap:disCenter}

We are using two different algorithms for wavefront correction in the experiment: GA and sGA. The implementation of sGA requires the optimization center (cx,cy) to be accurately aligned with the beam center (X, Y), as it serves as the center of symmetry for the algorithm. The method for determining the beam center (X, Y) on the SLM is discussed in Fig.~\ref{fig:2}. Here, we investigate how errors in locating (X, Y) impact the performance of both GA and sGA. To analyze this effect, the optimization center (the center of the applied phase front) is deliberately shifted by 5, 10, and 20 pixels, and the algorithms are executed to evaluate their effectiveness in wavefront correction. 

\begin{figure}[t]
\includegraphics[width=\linewidth]{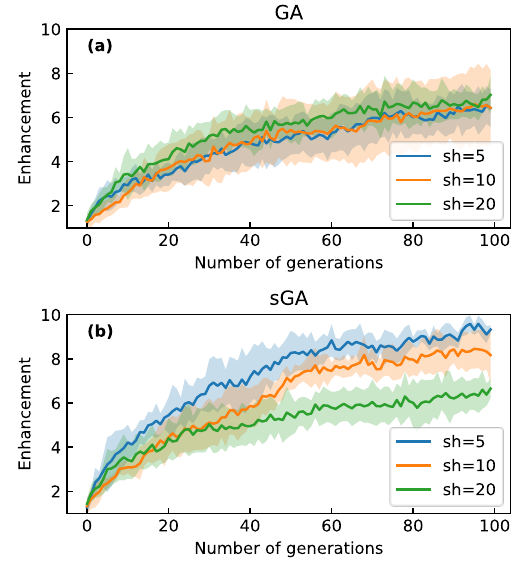}
\caption{\label{fig:S2} Effect of the displacement of the optimization center, (cx,cy) on the SLM relative to the beam center, (X,Y) for shifts of 5 pixels, 10 pixels, and 20 pixels, illustrated for (a) GA and (b) sGA. The disorder scale length of the diffuser is estimated to be approximately 32 pixels.}
\end{figure}

The results are presented in Fig.~\ref{fig:S2}, where (a) corresponds to GA and (b) to sGA, both tested under identical disorder realizations. 
The blue, orange, and green curves represent enhancements for shifts of 5, 10, and 20 pixels, respectively. Each data point is averaged over five independent runs with different random initial SLM configurations maintaining the same diffuser realization. The shaded regions represent one standard deviation of the statistical errors at each point, calculated across different runs. The findings show that while GA remains unaffected by shifts in the optimization center, sGA exhibits sensitivity to such displacements. This sensitivity arises because a shift misaligns the center of symmetry, causing the two entangled photons to experience different phase modulations from the SLM. As a result, the enhancement of coincidence counts decreases, with the reduction becoming more pronounced for larger shifts. The disorder scale length of the diffuser can be estimated from the correction phase mask shown in  Fig.~\ref{fig:3}(c).The phase undergoes a full $2\pi$ variation over approximately 4 super-pixels. Given that each super-pixel consists of an 8×8 pixel grid, the effective disorder scale length is around 32 pixels. The optimization center errors considered in this analysis represent a significant fraction of this scale length. The final enhancement decreases by 40$\%$ when the shift increases from 5 pixels to 20 pixels, which corresponds to approximately half the disorder scale length, for sGA. The rate of this decline is expected to depend on the disorder scale length, making the system more sensitive to errors in beam center determination for higher disorder strengths.

\begin{figure*}
\includegraphics[width=0.7\linewidth]{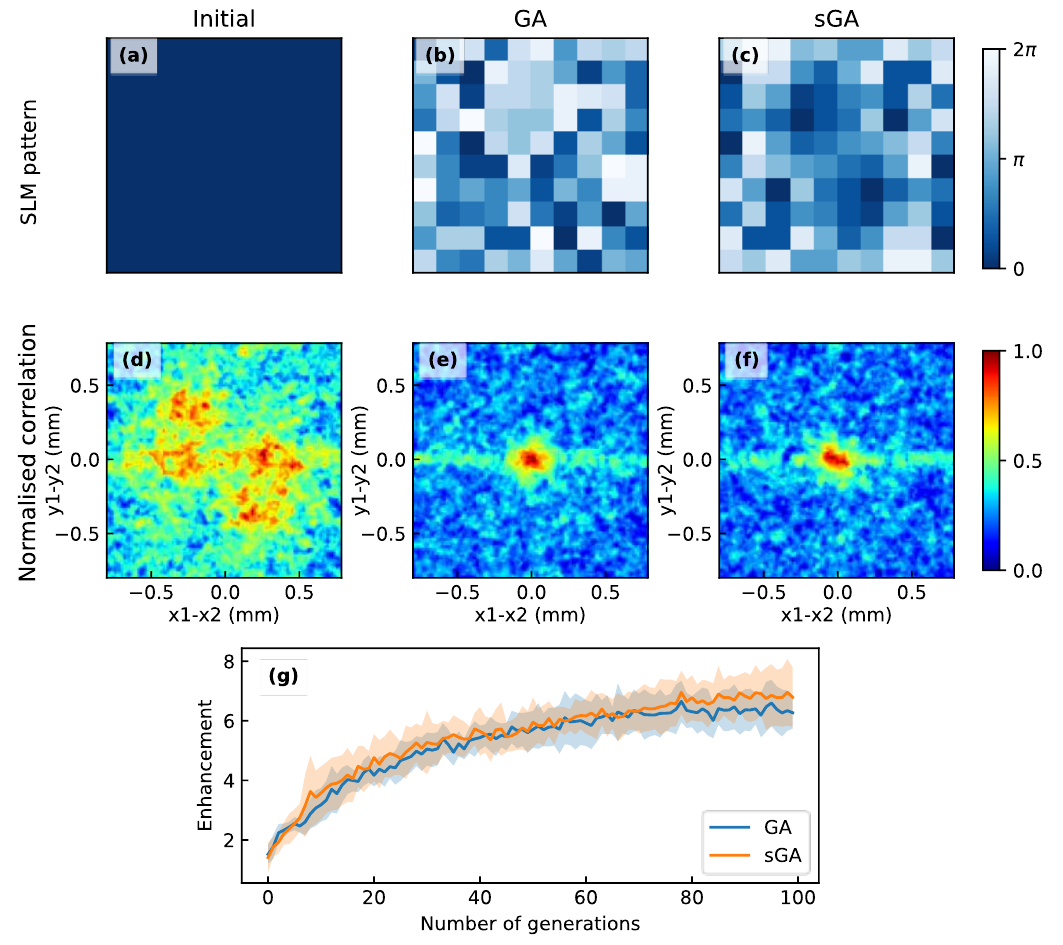}
\caption{\label{fig:S4}  Comparison of Wavefront Correction Algorithms when the detectors are placed at correlated points but not at the center of the beam. (a–c) SLM patterns before correction, after the genetic algorithm (GA), and after sGA (even part only). (d–f) Corresponding 2D correlations recorded via EMCCD. (g) Evolution of normalized coincidence counts at the correlated points over generations for GA and sGA.}
\end{figure*}

\section{Effect of displacement of detectors}\label{ap:ConSLM}
Here, we compare the performance of GA and sGA when the detectors are positioned at the correlated points but not at the beam center. For this comparison, both pinholes are deliberately displaced by one-fourth of the beam diameter in the detector plane in same direction. Fig.\ref{fig:S4} presents the results of this comparison. Panels (a)–(c) display the SLM pattern before correction, after correction using GA, and after correction using sGA, respectively. The corresponding 2D correlations obtained from EMCCD data are shown in panels (d)–(f). The contrast of the peaks shown in Fig.~\ref{fig:S4} (e) and (f) is 5.2 and 5.4 respectively. Fig.\ref{fig:S4} (g) tracks the enhancement evolution for both algorithms. In this scenario, both algorithms exhibit similar performance.

The degradation in the relative performance of sGA can be understood as follows: When the detectors are positioned away from the beam center, the single-photon counts and consequently, the coincidence counts decrease, weakening the feedback signal. As shown in Fig.~\ref{fig:4}, this reduced feedback signal strength affects both GA and sGA. The feedback strength can be improved by tilting the beam in the SLM plane, an odd function that sGA does not permit.However, GA allows this tilt function and can achieve better signal strength using that. Despite this limitation, both algorithms perform equally well in this case. However, as the detectors move further off-center, sGA's relative performance may degrade even more. This underscores the importance of precise detector positioning during optimization. The detectors can be correctly aligned by optimizing single-photon counts at each detector while adjusting the SLM tilt. Further, the use of a SPAD camera, which enables feedback from multiple correlated points simultaneously, can alleviate alignment sensitivity and improve overall feedback strength. 

\nocite{*}
\bibliography{aipsamp}

\end{document}